\documentstyle[a4,12pt,epsfig]{article}
\begin{document}
 
\begin{titlepage}
  \topmargin 1cm
   \begin{trivlist}
      \item[] \centerline{\Large{\bf{Plane Light-Like Shells and}}}
      \item[] \centerline{\Large{\bf{Impulsive Gravitational Waves}}}
      \item[] \centerline{\Large{\bf{in Scalar-Tensor Theories of Gravity}}}
      \itemsep 0.5cm
      \item[] \centerline{\normalsize{G.F. Bressange
		\footnote{bressang@celfi.phys.univ-tours.fr}}}
      \itemsep 0cm
      \item[] 	\begin{center}
		Laboratoire de Math\'{e}matiques et Physique Th\'{e}orique\\           
			UPRES-A 6083 du CNRS\\
      			Facult\'e des Sciences et Techniques\\
      			Parc de Grandmont 37200 Tours - France
		\end{center}
	\end{trivlist}
      \vspace{1cm}
      \begin{center}  
	\begin{trivlist}      
	\headsep    12cm 
	\item[]
	\centerline{\bf{Abstract}}   
	\itemsep   0cm  
	\item[] 
 
	We     study    gravitational   plane   impulsive  waves   and
	electromagnetic shock waves    in a scalar-tensor    theory of
	gravity  of the Brans-Dicke   type.  In vacuum, we  present an
	exact solution  of  Brans-Dicke's field equations and  give an
	example  in which a  plane  impulsive gravitational wave and a
	null shell of matter coexist on the same hypersurface.  In the
	homogenous case, we characterize them  by their surface energy
	density   and  wave  amplitude   and discuss the  inhomogenous
	case.  We also  give an  exact  solution of  the Brans-Dicke's
	field equations in the electrovacuum case  which admits a true
	curvature  singularity and use it  to built an example where a
	plane impulsive  gravitational  wave   and  an electromagnetic
	shock wave have the same null hypersurface as history of their
	wave fronts and propagate independently  and decoupled from  a
	null shell of   matter.   This  last  solution  is  shown   to
	correspond to the space-time describing the interaction region
	resulting  from the   collision of  two  electromagnetic shock
	waves leading to  the formation of two gravitational impulsive
	waves.  The  properties of  this  solution are  discussed  and
	compared to those  of  the Bell-Szekeres solution  of  general
	relativity. 

\end{trivlist} 
\end{center} 
\end{titlepage}
\newpage

%
%

\section{Introduction}
\topmargin -1cm 

	Nearly all the attempts at   deepening the connection  between
the  gravitational interaction and  the other interactions have led to
the  conclusion that  scalar fields  play   an important  role  in the
process of  quantization.  This idea first  came with the scalar field
(the     compacton) arising from  the  compactification   of the fifth
dimension in Kaluza-Klein theory \cite{KK}.  Later, the existence of a
scalar field   found its justification     in the  Brans-Dicke   model
\cite{FJBD}  and  its further generalizations including  in particular
non-linear   $\sigma$-models  with  several  scalar fields \cite{BDG}.
More recently, the low energy limit of string theory has been found to
exhibit a scalar field coupled to gravity, the dilaton, having a local
space-time dependent  coupling  to the matter  fields \cite{GSW} while
the  scalar fields in scalar-tensor  theories  of the Brans-Dicke type
present a universal metric coupling to matter \cite{BB}. Scalar fields
are also believed  to be important in  cosmology, for  instance in the
scenarios of inflation (inflaton) and extended inflation (inflaton and
Brans-Dicke field) \cite{St}. 

	Plane wave geometries have many interesting properties in four
dimensions  such as being  geodesically complete and admitting metrics
belonging to the family of  the  pp-waves \cite{EK}, \cite{JBG}  which
are  plane-fronted waves with parallel rays.   They  are known since a
long time to be exact  solutions of Einstein's vacuum field equations.
More recently,    in heterotic  string theory,  by   investigating the
possible modifications   in   superstring effective  theories   of the
$d=10$, $N=1$   Einstein-Yang-Mills supergravity  theory, it has  been
shown that these plane geometries are  exact solutions at all order of
the string   tension  parameter \cite{Guv}.   Later,  exact  solutions
describing colliding impulsive gravitational  waves in dilaton gravity
have been obtained \cite{GS}  ( these solutions have the Bell-Szekeres
solution of general relativity \cite{BS} as a limiting case) and other
ones  have been derived by  using  the method  of the harmonic mapping
combined    with    the    algebra   associated    with     the  group
$SL(2,\mbox{I\hspace{-.15em}R})$ \cite{BMG}. 

	In  this   paper, we describe  some   plane geometric features
appearing in a scalar tensor theory of gravity such as the Brans-Dicke
theory and  consider    also   the  coupling    of  this  theory    to
electromagnetism.   More precisely, the   purpose of this  work is  to
study impulsive gravitational  waves  conjointly with null shells   of
matter and electromagnetic shock  waves.    In general relativity,   a
solution describing the  collision of a scalar  plane wave with either
another  scalar plane wave,  or an  impulsive  gravitational wave or a
neutrino wave has been obtained \cite{WU}.   On the other hand, planar
domain walls solutions coupled  to   the Brans-Dicke field have   been
given \cite{SW} \cite{BB}.  Here, we  describe situations in which  an
impulsive gravitational wave, a null  shell of matter and possibly  an
electromagnetic shock wave coexist  on the same null hypersurface. For
that purpose, we  use a formalism  recently developped \cite{BB} which
provide, in  the  context  of scalar-tensor  theories  of  gravity,  a
unified description of all types of hypersurfaces (timelike, spacelike
and  lightlike) and  allows  the space-time  coordinates  to be chosen
freely and independently on both sides of  the shell in agreement with
their own symmetries. 

	The paper is organized as follows. In section 2, we review the
shell  formalism and summarize  the    principal results obtained   in
\cite{BB}  for the null case in  the Brans-Dicke theory  and also some
additional results obtained in \cite{BBH} about the general separation
between a null shell and a wave.  In section  3, the Brans-Dicke vacuum
case is  considered.   In the   context of plane  symmetry,  an  exact
solution of the Brans-Dicke field equations  is presented and is glued
to  the Minkowski space-time along     a  null hypersurface.  This
re-attachement is accomplished  by making a suitable identification of
the coordinates on  both sides of  the shell in  terms of an arbitrary
shift function. A null shell of  matter and an impulsive gravitational
wave are  shown to coexist on  this null hypersurface and  the surface
energy density  of the shell and the  wave amplitude are determined in
the  homogenous case.  The  inhomogenous  case is  also discussed.  In
section  4, the electrovacuum    case is considered.   An  exact plane
symmetric   solution  of the   Brans-Dicke's   field equations  in the
presence of a Maxwell field is given and it is found to present a true
curvature    singularity.  This solution    is glued  to the vacuum solution 
of section 3 along a null hypersurface and represents the history of the
wave  fronts    of a   gravitational    impulsive  wave   and   of  an
electromagnetic shock wave propagating with  and decoupled from a null
shell of matter carrying no  surface-current.  Finally, it is remarked
that  the given exact solution can  be interpreted  as the interaction
resulting from the   collision of two  electromagnetic  shock waves of
constant  aligned   polarization propagating   in a  conformally  flat
background and   approaching from    opposite directions.   After  the
collision, two impulsive gravitational waves  are produced and coexist
with the  original  electromagnetic  shock waves.  This   provides the
analog  in  the Brans-Dicke theory  of the   Bell-Szekeres solution of
general relativity \cite{BS} which is known to be diffeomorphic to the
Bertotti-Robinson  space-time  \cite{BR} \cite{LC}.  A difference arises however
because  the    interaction region  is  not    conformally flat  as  a
consequence  of  the  scattering  of   the  gravitational  waves   and
furthermore because of the presence of a true curvature singularity. 

%
%

\section{General formalism for a null shell in Brans-Dicke theory}

	In  this   section are summarized the   main   properties of a
singular null hypersurface in  the   Brans-Dicke theory.  A   complete
description of an arbitrary singular hypersurface (timelike, spacelike
or lightlike)  in the scalar-tensor theories of  gravity, of which the
Brans-Dicke theory is a particular case, having already presented in a
recent paper  \cite{BB}, we  only recall   the results  which  will be
useful for the present work and refer the reader to \cite{BB} for more
details.   Furthermore, more  information on the  justification of the
separation between a shell part and a wave part which generally occurs
on a lightlike hypersurface can be found in \cite{BBH}. 

	In   the Brans-Dicke   theory,  it  is    well-known that  two
conformally  related metrics  can be used   :  the Jordan-Fierz metric
usually referred to as the physical metric  (because the matter fields
are minimally coupled to it  and the stress-energy tensor is conserved
relatively to it), and the  Einstein metric, where an Einstein-Hilbert
term is  recovered in    the action   (in   the Einstein   frame,  the
stress-energy tensor  of the matter fields  is not  conserved).  These
frames  provide  two equivalent descriptions and  as  in \cite{BB}, we
shall work in the Einstein-frame where the equations of a shell take a
simpler form. The space-time $M$ in the  Brans-Dicke theory is endowed
with  a pair $(g_{\mu   \nu},\varphi)$   where $g_{\mu \nu}$ is    the
space-time metric and  $\varphi$ a scalar  field.  In the  presence of
matter represented by a stress-energy tensor  $T_{\mu \nu}$, the field
equations  for the  pair $(g_{\mu \nu},\varphi)$  are  in the Einstein
frame 
\begin{eqnarray}
R_{\mu \nu}&=&\,8\pi\,G       \,\left(T_{\mu  \nu}     -{T\over     2}
			g_{\mu\nu}\right)  +2\,{\partial}_{\mu}\varphi
			{\partial}_{\nu}\varphi \\ \Box  \varphi   &=&
			-4\pi G\,\alpha T 
\end{eqnarray}
where  $\Box$ stands  for the usual  d'Alambertian operator associated
with the metric  $g_{\mu \nu}$ and  where $T$ represents the trace  of
$T_{\mu \nu}$ with respect to $g_{\mu \nu}$.   In (2), $\alpha$ is the
constant coupling  factor of the scalar  field $\varphi$ to matter and
it is related to the  usual Brans-Dicke parameter $\omega$ by  $\alpha
^2={(2\omega+3)}^{-1}$.  In order to get  the physical or Jordan-Fierz
metric   $\tilde{g}_{\mu\nu}$, one simply   has  to  do the  conformal
transformation $\tilde{g}_{\mu\nu}=e^{\alpha\varphi}g_{\mu\nu}$. 

	Let us  consider  a null hypersurface  $\Sigma$ resulting from
the isometric soldering  of  two isometric null hypersurfaces  $\Sigma
^+$ and $\Sigma ^-$ respectively embedded in the space-times $M^+$ and
$M^-$,  and further  identify $\Sigma$ as  a null  hypersurface in the
manifold  $M^+  \cup  M^-$.   Let us  introduce  two  local systems of
coordinates,   $\left\lbrace  x^{\mu}_-  \right\rbrace$  in  $M^-$ and
$\left\lbrace x^{\mu}_+ \right\rbrace$  in  $M^+$, the  indices  $\mu$
running from $0$ to $3$.   Relatively to those systems of coordinates,
we denote by $g_{\mu \nu}^+$ and $g_{\mu \nu}^-$ the components of the
metric tensor , by $\varphi ^+$ and $\varphi ^-$ the scalar field, and
by $T_{\mu \nu}^+$  and $T_{\mu \nu}^-$ the  components of  the matter
stress-energy  tensors  in  each domain $M^+$   and  $M^-$.  The field
equations (1)-(2) hold  separetly in $M^+$ and $M^-$  for the two sets
$(g_{\mu \nu}^+,\varphi ^+,T_{\mu \nu}^+)$ and $(g_{\mu \nu}^-,\varphi
^-,T_{\mu \nu}^-)$ respectively. In the present work, $\Sigma$ 
represents the history of a shell and/or
an impulsive wave, and the metric and  the scalar field are only $C^0$
on            it    i.e.        $[g_{\mu\nu}]=[\varphi]=0$         but
$[\partial_{\alpha}g_{\mu\nu}]\neq               0$                and
$[\partial_{\alpha}\varphi]\neq 0$ where the brackets denote the jump 
across $\Sigma$ of the quantity contained therein. 

	Let us  introduce a  set of  intrinsic  coordinates 
$\xi ^a$ ($a\,=\,1,2,3$ ) on $\Sigma$, the corresponding tangent basis
vectors $e_{(a)}={\partial \over  {\partial \xi^a}}$ and the common 
induced metric $g_{ab}$ on $\Sigma$ from $\,M^+$ and $M^-$. In  the case of  a 
lightlike hypersurface,  the  normal $n$ is
null and tangent  to the  surface.   In  order  to  describe how the
hypersurface  is  embedded in   spacetime,   we need  to   introduce a
transverse vector $N$ on $\Sigma$ such that 
\begin{equation}
N.n\,=\,{\eta}^{-1}\hspace{1cm}, 
\end{equation}
where $\eta$ is  a  given non-vanishing  smooth function on  $\Sigma$.
This transversal  is  uniquely  defined  by (3)  up   to  a tangential
displacement.  In order to make sure  that we consider the same normal
and transversal on both sides, we impose the relations 
\begin{equation} \left\lbrack n.e_{(a)} \right\rbrack\,=\,
\left\lbrack    N.e_{(a)}   \right\rbrack\,=\,     \left\lbrack    N.N
\right\rbrack\,=\,0\hspace{1cm}.  \end{equation}    We      define the
transverse extrinsic curvature  of $\Sigma$ on   the $+$ side  and $-$
side by 
\begin{equation}
{\cal
K}_{ab}^{\pm}\,=\,-\,N.{\nabla}_{e_{(b)}}e_{(a)}{\vert}^{\pm}\hspace{1cm}.
\end{equation}
Because the metric is only continuous  across $\Sigma$, the transverse
extrinsic curvature is discontinuous across $\Sigma$ and its jumps are
denoted 
\begin{equation}
{\gamma}_{ab}\,=\,2\,\left\lbrack    {\cal  K}_{ab}      \right\rbrack
		\hspace{1cm}. 
\end{equation}
It can be shown that  the $3$-tensor ${\gamma}_{ab}$'s is  independent
on the  choice of the  transverse  vector field  $N$  and provides  an
intrinsic description of  the embedding of  ${\Sigma}$ in the manifold
$M^+\cup M^-$.  It can   be extended  (in  a  non unique way)  into  a
$4$-tensor ${\gamma}_{\mu \nu}$ in either $M^+$  or $M^-$ by requiring
its   projection   on    ${\Sigma}$   to be     ${\gamma}_{ab}$   i.e.
$\,{\gamma}_{\mu \nu}e_{(a)}^{\mu}e_{(b)}^{\nu}\,=\,{\gamma}_{ab}$. 

	Introducing  these  results into  the field equations (1)-(2),
one can show that the matter stress-energy tensor $T^{\mu \nu}$ admits
a  singular Dirac  part $S^{\mu  \nu}$ concentrated  on ${\Sigma}$ and
takes the following general form 
\begin{equation}
	T^{\mu        \nu}\,=\,\chi\,S^{\mu       \nu}\,{\delta}(\Phi)
	+T^{\mu\nu}_{\pm}\hspace{1cm}, 
\end{equation}
where  $\Phi(x^{\mu})=0$ is  the equation   of ${\Sigma}$  in  a local
system of coordinates  and $\chi$ is  a normalization factor such that
$n\,=\,{\chi}^{-1}\,{\nabla}\Phi$.  The surface   stress-energy tensor
$S^{\mu \nu}$ is given by \cite{BB} 
\begin{equation}
16\pi\,{\eta}^{-1}\,S^{\mu
				\nu}\,=\,2\,{\gamma}^{(\mu}n^{\nu)}-{\gamma}\,n^{\mu}
				n^{\nu} \hspace{1cm}, 
\end{equation}
with 
\begin{equation}
{\gamma}^{\mu}={\gamma}^{\mu \nu}n_{\nu}\,\,;\,\, \gamma   =    g^{\mu
		\nu}{\gamma}_{\mu \nu}\hspace{1cm}, 
\end{equation}
These quantities can  be calculated on either side  of $\Sigma$ and it
can be proven that the expression (8) is independent of the extension
from   ${\gamma}_{ab}$ to ${\gamma}_{\mu   \nu}$.  The  surface-stress
energy   tensor   $S^{\mu   \nu}$   is    purely  tangential,  $S^{\mu
\nu}n_{\nu}=0$,  and  it can  also  be expressed in  the tangent basis
$e_{(a)}$ as an intrinsic $3$-tensor $S^{ab}$ given by 
\begin{equation}
16\pi      G\,      {\eta}^{-1}\,S^{ab}\,=\left(g_{*}^{ac}\,l^{b}l^{d}
		+l^{a}l^{c}\,g_{*}^{bd}-l^{a}l^{b}\,g_{*}^{cd}
		\right){\gamma}_{cd} \hspace{1cm} 
\end{equation} 
where the $l^a$'s are such that 
\begin{equation}
		n\,=\,l^a e_{(a)} \hspace{1cm}, 
\end{equation}
and satisfy $g_{ab}l^b = 0$.  The quantity $g_{*}^{ab}$ is defined by 
\begin{equation}
		g_{*}^{ab}\,g_{bc}^*={\delta}^a_c-\eta l^a
		N_c\hspace{1cm}. 
\end{equation}

	The Weyl tensor of the global space-time $M^+ \cup M^-$ has in
general a  Dirac part concentrated  on  $\Sigma$ which  is given as in
general relativity \cite{BI} by 
\begin{equation}
C^{\alpha \beta}_{\ \ \mu \nu} =  \lbrace \,2 \eta\, n^{\lbrack\alpha}
{\gamma}^{\beta\rbrack}_{\lbrack\mu}     n_{\nu\rbrack}  -  16 \pi  \,
{\delta}^{\lbrack\alpha}_{\lbrack\mu} S^{\beta\rbrack}_{\nu\rbrack}  +
\frac{8}{3} \pi  \,S^{\lambda}_{\lambda}  {\delta}^{\alpha \beta}_{\mu
\nu}\,\rbrace \,\chi\, \delta(\Phi) \hspace{1cm}. 
\end{equation}
An important  result  is that   the ${\gamma}_{ab}$'s split  into  two
independent  parts, which are   separetly associated with  a shell and
with a   wave.  It is easy   to see from (8)  or  (10) that  only the
components  ${\gamma}_{ab}l^b$  and  $\gamma=g^{\mu  \nu}{\gamma}_{\mu
\nu}=g_{*}^{cd}{\gamma}_{cd}$  of the ${\gamma}_{ab}$'s contribute  to
the expression  of    the stress-energy   tensor.   This  leaves   two
independent components denoted  by $\hat{\gamma}_{ab}$ representing an
impulsive gravitational wave and related to the two degrees of freedom
of  polarization    of  the      wave.   The    expression   of    the
$\hat{\gamma}_{ab}$'s is \cite{BB} \cite{BBH} 
\begin{equation}
 \hat{\gamma}_{ab} = \gamma_{ab}  -  \frac{\gamma}{2} g_{ab} + \eta  (
	N_{a} \gamma_{bc} + N_{b} \gamma_{ac} ) l^c\hspace{1cm}, 
\end{equation}
The impulsive gravitational  wave   propagates with the shell   in the
direction $n^{\mu}$ in space-time and with $\Sigma$  as history of its
wave-fronts. 

	In the case when  an electromagnetic field ${F}_{\mu \nu}$  is
present,   it may     be  discontinuous   across   $\Sigma$  and   the
discontinuities   of the  field are   related  to the existence of  an
electromagnetic shock wave and  to surface-currents (note that this is
also   valid  for an arbitrary    gauge  field  -see \cite{BB}).   The
electromagnetic potential $A_{\mu}$ is continuous across $\Sigma$ {\it
i.e.}  $\left\lbrack A_{\mu} \right\rbrack \,=\,0$, but its transverse
derivatives are not and one introduces a vector ${\lambda}^{\nu}$ such
that 
\begin{equation}
	[{\partial}_{\mu}           A_{\nu}]=\eta              n_{\mu}
	{\lambda}_{\nu}\hspace{1cm}. 
\end{equation}
The jump of the electromagnetic field across $\Sigma$ is therefore 
\begin{equation}
	[{F}_{\mu   \nu}]=\eta   (  n_{\mu}  {\lambda}_{\nu}-  n_{\nu}
				 {\lambda}_{\mu}) \hspace{1cm}. 
\end{equation}
From the Maxwell equations, it can be shown that there exist a surface
current $j^{\mu}$ such that 
\begin{equation} 
	4\pi j^{\mu} \ =\ \eta (\lambda .n) n^{\mu} \hspace{1cm}. 
\end{equation}
This expression shows that  $j^{\mu}$ is purely tangential to $\Sigma$
and  that only  the    part  $\,{\lambda}.n\,$  of   ${\lambda}_{\mu}$
contributes to the surface current. The remaining part 
\begin{equation}
     \hat{\lambda}             ^{\mu}           \,=\,          \lambda
^{\mu}-\eta(\lambda.n)\,N^{\mu}\hspace{1cm}, 
\end{equation}
characterizes the electromagnetic shock wave. 
	
%
%

\section{Pure vacuum}


	In   this section,  we    first solve  the  Brans-Dicke's  field
equations  (1)-(2) in the   case of vacuum  ($T^{\mu\nu}=0$) and plane
symmetry.  Then we apply  our solution to the  description of a  plane
null shell and impulsive gravitational wave.  It is convenient to take
the metric in the form given by Szekeres \cite{S} 
\begin{equation}
 ds^2 = -2 e^{-M}dudv+e^{-U} (e^V dx^2 + e^{-V} dy^2)\hspace{1cm}, 
\end{equation}
where the functions $M$, $U$, $V$ and the  scalar field $\varphi$ only
depend  on the null coordinates  $(u,v)$,   and $(x,y)$ are  spacelike
coordinates  - we use  the ordering $u,v,x,y$  and greek indices range
from  $0$ to $3$.   The form of the  field equations  (1)-(2) with the
metric (19) and   an arbitrary  stress-energy tensor $T^{\mu\nu}$   is
given  is appendix   A.  The space-time  defined   by the  metric (19)
contains   plane  waves   propagating  along  the  null  hypersurfaces
$u=constant$  and $v=constant$  as demonstrated by   Bondi, Pirani and
Robinson \cite{BPR}\cite{JBG}. 

	In vacuum, the field equations  reduce to the following set of
	equations 
\begin{eqnarray}
2\,U_{uu}-{U_u}^2-V{_u}^2+2\,M_u\,U_u\,=\,4\,{\varphi          _u}^2\\
2\,U_{vv}-{U_v}^2-V{_v}^2+2\,M_v\,U_v\,=\,4\,{\varphi          _v}^2\\
U_u\,U_v-U_{uv}\,=\,0\\            2\,V_{uv}-U_u\,V_v-U_v\,V_u\,=\,0\\
U_{uv}+2\,M_{uv}-V_u\,V_v\,=\,4\,\varphi _u\,\varphi   _v\\ 2\,\varphi
_{uv}-U_u\,\varphi _v-U_v\,\varphi_u\,=\,0 
\end{eqnarray} 
where  the indices $u$  and  $v$  refer  to partial derivatives   with
respect  to   these  variables.   Equation  (22) can    immediately be
integrated as 
\begin{equation} 
e^{-U}\,=\,f(u)+g(v) 
\end{equation} 
where $f$ and $g$
are arbitrary functions and,  following Szekeres \cite{S}, the general
solution of (23) is 
\begin{equation}
	V={\lambda}_1 \tanh  ^{-1}   \left\lbrack   \sqrt{{1\over 2}-f
			\over  {1\over 2}+g}\right\rbrack +{\lambda}_2
			\tanh   ^{-1} \left\lbrack \sqrt{{1\over  2}-g
			\over {1\over 2}+f}\right\rbrack 
\end{equation}
where  ${\lambda}_1$  and  ${\lambda}_2$ are two  arbitrary constants.
Moreover,  as the equation  (25)  for the   scalar field  $\varphi$ is
similar to (23), one gets for the scalar field \cite{WU} 
\begin{equation}
	\varphi      \,=\,{\lambda}_3   \tanh   ^{-1}     \left\lbrack
			\sqrt{{1\over      2}-f     \over      {1\over
			2}+g}\right\rbrack +{\lambda}_4  \tanh   ^{-1}
			\left\lbrack \sqrt{{1\over 2}-g  \over {1\over
			2}+f}\right\rbrack +{\varphi}_0 
\end{equation}
where ${\lambda}_3$,    ${\lambda}_4$  and  $\varphi   _0$  are  three
arbitrary constants.   In order for  (26-28) to represent well-defined
solutions, the functions $f$ and $g$ must satisfy $0<f<{1\over 2}$ and
$0<g<{1\over 2}$.  Equation (20) then gives 
\begin{equation}
M_u\,=\,-{f^{''}(u)\over f^{'}(u)} -{f^{'}(u)\over 2(f+g)}\left\lbrack
				\alpha   ^2 \left({{1\over   2}+g\over
				{1\over  2}-f}\right)  +\beta       ^2
				\left({{1\over    2}-g\over    {1\over
				2}+f}\right) +\gamma    {\sqrt{{1\over
				2}+g}\sqrt{{1\over    2}-g}      \over
				\sqrt{{1\over       2}+f}\sqrt{{1\over
				2}-f}} -1 \right\rbrack 
\end{equation}
where we have introduced the scalars 
\begin{equation}
	\alpha^{2}\,=\,{\lambda_1^2\over       4}+\lambda_3^2\,\,,\,\,
	 \beta^{2}\,=\,{\lambda_2^2\over   4}+\lambda_4^2    \,\,,\,\,
	 \gamma\,=\,{\lambda_1\lambda_2\over 4}+\lambda_3\lambda_4 
\end{equation}
A similar expression can  be obtained for  $M_v$ from equation (21) by
interchanging $f(u)$   with $g(v)$ and  $\alpha$  with $\beta$.  Then,
integrating $M_u$ and $M_v$, one gets 
\begin{eqnarray*}
M &  = & -\log(k  f^{'}g^{'})-{\alpha  ^{2}+\beta ^2+2\,\gamma-1 \over
2}\log(f+g)+  {\alpha^{2}\over   2}\log\left(({1\over    2}-f)({1\over
2}+g)\right)\\  & &\hspace{2cm}+{\beta^{2}\over   2}\log\left(({1\over
2}+f)({1\over 2}-g)\right) +2\,\gamma\,\log D 
\end{eqnarray*}
where  $D\,=\,\sqrt{{1\over     2}-f}\sqrt{{1\over 2}-g}+\sqrt{{1\over
2}+f}\sqrt{{1\over 2}+g}$ and   $k$ is an  arbitrary  constant.   This
expression is similar to  general relativity -see \cite{JBG}- but  the
constants $\alpha$, $\beta$   and $\gamma$ now  include those  of  the
Brans-Dicke scalar field. 

It   can be checked that the   remaining  equation (24) is identically
satisfied when $U$, $V$, $M$  and $\varphi$ are  replaced by the above
expressions.   Hence, these  expressions  constitute a plane symmetric
solution of the vacuum Brans-Dicke field equations.  As we shall later
be concerned with a  plane wave or a null  shell along $u=const.$,  we
now consider   the case where the   metric and the  scalar  field only
depend  on $u$.   The  corresponding solution  is obtained by  putting
$g={1\over    2}$ in   (27)       and   (28)  and   taking     $\alpha
^{2}\,=\,{\lambda_1^2\over 4}+\lambda_3^2$, and $\beta=\gamma=0$.  One
then gets 
\begin{eqnarray}
    e^{-U} &  = &  f+{1\over 2}\\  e^V  & =  & {\left({1+\sqrt{{1\over
2}-f}\over  1-\sqrt{{1\over  2}-f}}\right)}  ^{\lambda  _1  \over 2}\\
e^{-M}    &   =  & k    f^{'}(u){{\left({1\over  2}+f\right)}^{{1\over
2}\left(\alpha ^2-1\right)} \over  {\left({1\over 2}-f\right)}^{\alpha
^2}}\\ e^{\varphi} &  = & e^{\varphi _0} {\left({1+\sqrt{{1\over 2}-f}
\over   1-\sqrt{{1\over     2}-f}}\right)}      ^{\lambda  _3    \over
2}\hspace{2cm}. 
\end{eqnarray}


	Let us now consider two space-times  $M^+$ and $M^-$ separated
by  a null hypersurface $\Sigma$   located at $u=0$.  We suppose  that
$M^+$  has  coordinates $(u,v^+,x^+,y^+)$ and  a  metric and  a scalar
field  given by (31-34).   $M^-$    is the Minkowski space-time   with
coordinates $(u,v,x,y)$,  a metric given by (19)  with $M=U=V=0$ and a
constant  scalar field  (we drop the  minus   indices on  any quantity
belonging to $M^-$).  A Brans-Dicke solution can be re-attached to the
Minkowski space-time  only if the scalar field  takes this same constant
value on the surface of junction.  The two space-times are glued along
the null hypersurface $\Sigma$, $u=0$, by making the identification 
\begin{equation}
	(0, v^+, x^+, y^+)\,=\,(0, v-F(x,y), x, y) 
\end{equation}
where $F$ is   an arbitrary function  of  the coordinates $x$  and $y$
producing    a  shift in   the  null  coordinate    $v$ tangent to the
hypersurface. This identification was previously introduced by Penrose
to  generate  a plane impulsive gravitational  wave   in the Minkowski
space-time \cite{Pen}. 

	The  condition of  continuity  of the  metric  across $\Sigma$
requires that 
\begin{equation}
	f(0)\,=\,{1\over 2} \hspace{1cm} 
\end{equation}
In order for $M$ expressed in (33)  to be defined  at $u=0$, we choose
the arbitrary function $f(u)$ such that 
\begin{equation}
	f(u)\,=\,{1\over 2}-(\mu u)^{\rho},\hspace{1cm}\rho\,>\,0 
\end{equation}
with 
\begin{equation}
	\rho   \left(1-{\alpha^2  \over  2}\right)\,=\,1\,\,,\,\,1\leq
 \alpha^2<2\,\,. 
\end{equation} 
Although  no condition of continuity need  to be imposed  on $M$ as it
does not contribute to the induced metric on $\Sigma$, one chooses the
arbitrary constant $k=-(\mu \rho)^{-1}$ and obtains for $M$ 
\begin{equation}
	e^{-M}\,=\,(1-(\mu u)^{\rho})^{\alpha^2-1\over 2} 
\end{equation}
making   $M$ equal to its  Minkowski  value $M=0$  at $u=0$.  Finally,
using  (34),  the continuity  of the scalar   field across $\Sigma$ is
automatically satisfied provided that the constant value of the scalar
field in $M^-$ is ${\varphi}_0$. 

	We take  $\xi^a =  (v,x,y)$  with  $a  = 1,2,3$,  as intrinsic
parameters on $\Sigma$, and choose the normal to be $n = e_{(1)}$.  We
obtain from (11) $l^a =  \delta^a_1$.  The induced metric is $\,g_{ab}
= diag(0,1,1)$ and one may take  for its 'inverse' (12), $\,g^{ab}_* =
diag(0,1,1)$.  A convenient choice for the transversal $N$ corresponds
to $N.n = -1$, $N.e_{(2)} = N.e_{(3)} = 0$, $N.N = 0$, thus leading to
components equal to $N^\alpha = (1,0,0,0)$ in $M^-$, and 
\begin{equation}
N^{\alpha}_+  =  \left(1,   \frac{F^2_x  +  F^2_y}{2},    -F_x,   -F_y
\right)\hspace{0.5cm}, 
\end{equation}
in  $M^+$.    Introducing these  results  into (5),  one   obtains the
transverse extrinsic  curvature ${\cal K}_{ab}^{\pm}$  on each side of
the shell.   As  ${\cal  K}_{ab}^-$  vanishes,  one  gets  from   (6),
$\gamma_{ab} = 2{\cal  K}^{+}_{ab}$   and the non-zero  components  of
$\gamma_{ab}$ are equal to 
\begin{eqnarray}
\nonumber  & & \gamma_{22}  = -2F_{xx}\\ &  & \gamma_{33} = -2F_{yy}\\
	  \nonumber    &     &    \gamma_{23}   =    \gamma_{32}     =
	  -2F_{xy}\hspace{1cm}. 
\end{eqnarray}
Therefore  the surface stress-energy  tensor  (10) is of  the form, $-
S^{ab} =  \sigma l^a l^b $, and  has only one  non-vanishing component
equal to 
\begin{equation}
  16 \pi G S^{11} = \gamma_{22} + \gamma_{33} = -2\,\Delta F 
\end{equation}
where $\Delta$  is  the Laplacian  in the  $2$-dimensional  flat space
$(x,y)$.  These results show that  the shell has  no shear and is only
characterized by a surface energy density $\sigma$ equal to 
\begin{equation}
     \sigma = {\Delta F\over 8 \pi G } \hspace{1cm}. 
\end{equation}
which  requires $\Delta F  \geq  0$ to ensure   the positivity  of the
energy density. 

	On the other hand the wave  part (14) of $\gamma_{ab}$ has the
only non-vanishing components 
\begin{eqnarray}
&  &     \hat\gamma_{22}      =    -\hat\gamma_{33}       =    -F_{xx}
+F_{yy}\hspace{1cm},\\ & & \hat\gamma_{23} = -2F_{xy}\hspace{1cm}. 
\end{eqnarray}
	
	The above  results show  that  a null  shell and an  impulsive
gravitational  wave generally  coexist and  that  their properties are
characterized by the shift function $F(x,y)$.   The shell and the wave
propagate independently and  have the same null hypersurface  $\Sigma$
as world sheet. The homogeneous case is obtained by taking 
\begin{equation}
	F(x,y)={A\over  4}(x^2+y^2)-{B\over 4}(x^2-y^2)-{C\over 2} x y
	\hspace{1cm} 
\end{equation}
where $A$, $B$ and $C$ are three constants with  $A\geq 0$.  The shell
has a constant  energy density such that $8\pi  G \sigma = A$  and for
the      wave       we  have  $\hat{\gamma}_{22}=\hat{\gamma}_{33}=B$,
$\hat{\gamma}_{23}=C$. If $A=0$, we have only a wave  and only a shell
if $B=C=0$.  Inhomogenous  plane shells and waves  are obtained if the
shift  function $F(x,y)$ is arbitrary.  The expansion $\rho$ and shear
$\sigma$ for the null geodesics with tangent  vector $N$ transverse to
the hypersurface  $\Sigma$ vanish on the $M^-$  side, and on the $M^+$
side, they are equal to 
\begin{eqnarray}
	\rho^+   & = & N_{\mu;\nu}m^{\mu}\bar{m}^{\nu}\,=\,-{1\over 2}
	(F_{xx}+F_{yy})=-4\pi  G     \sigma\\  \sigma^+     &    =   &
	N_{\mu;\nu}m^{\mu}m^{\nu}\,=\,-{1\over                      2}
	(F_{xx}-F_{yy})-iF_{xy}={1\over     2}    (\hat{\gamma}_{22}+i
	\hat{\gamma}_{23}) 
\end{eqnarray}
where    we    have used  the  null     tetrad   $(n,N,m,\bar m)$ with
$m\,=\,(e_{(2)}+ie_{(3)}){/\sqrt 2}$   and where $;$   stands for  the
covariant derivative.  Since $\rho^+\leq 0$, the null
congruence  is focussed after  having crossed the   shell with energy  density
$\sigma\geq 0$ and  from (48), one sees that the wave part is  responsible for the shear of
the null geodesics. 

	In this paper, we have  glued the two space-times $M^{\pm}$ by
making the identification (35)  and have found  that the shell is only
characterized by its surface-energy density. When the relation between
the  coordinates $v^+,x^+,y^+$ and $v,x,y$  takes a more general form,
we expect the  shell to admit  also a surface pressure and anisotropic
surface stresses.

%
%

\section{Electrovacuum}


	In this section, we first present an exact plane wave solution
of   the  Brans-Dicke   field   equations in     the  presence of   an
electromagnetic field;  this solution admits  a curvature singularity.
Then, using this solution, one builts an example  of a plane impulsive
gravitational wave propagating  together  with a null shell  of matter
and an electromagnetic   shock  wave on the   same  null hypersurface.
Finally,  one shows that our exact  solution describes the
interaction region resulting from the collision of two electromagnetic
shock waves propagating in a  Brans-Dicke conformally flat background.
This provides  the analogue of  the  Bell-Szekeres solution in general
relativity  \cite{BS}  but  the   properties  obtained  here  in   the
interaction region are completly different because of  a presence of a
curvature singularity. 
 
	As the  electromagnetic stress-energy tensor  $T_{\mu \nu}$ is
traceless, the Brans-Dicke's field equations reduce to 
\begin{eqnarray}
R_{\mu \nu}&=&\,8\pi\,G \,T_{\mu     \nu}  +2\,{\partial}_{\mu}\varphi
		\,{\partial}_{\nu}\varphi \\    \Box \varphi     &=& 0
		\hspace{1cm} 
\end{eqnarray}
and  they have  to be solved  simultanously  with  the  Maxwell vacuum
equations 
\begin{equation}
		\nabla_{\mu}F^{\mu\nu}=0 
\end{equation}
We still use the Szekeres form (19) of the metric as we still have the
plane symmetry and we take for the  electromagnetic potential the 
form recently introduced in \cite{Hogan} to describe plane gravitational 
waves in Bertotti-Robinson spacetimes
\cite{BR} \cite {LC} \cite{EX} 
\begin{equation}
A\,=\,\cos\theta\,\sin    (au-bv)\,dx+\sin\theta   \,\sin  (au+bv)\,dy
		\hspace{1cm}. 
\end{equation}
where $a$, $b$ and $\theta$ are  three arbitrary constants.  Using the
expressions (A.7-A.10) for the components of $T_{\mu \nu}$ in terms of
the metric functions and the  field equations (A.1-A.6), we obtain the
following solution 
\begin{equation}
U\,=\,M\,=-\log   \left\lbrack  {\cos    (au-bv)   \, \cos    (au+bv)}
\right\rbrack\hspace{1cm}, 
\end{equation}
\begin{equation}
V\,=\,\log   \left\lbrack     {\cos  (au-bv)  \over    \cos   (au+bv)}
\right\rbrack\hspace{1cm}, 
\end{equation}
\begin{equation}
\varphi \, = \,{1\over  \sqrt 2}\, \log  \left\lbrack {\cos (au-bv) \,
	\cos     (au+bv)}   \right\rbrack\,      +\,       {\varphi}_0
	\,\,,\,\,{\varphi}_0=const\hspace{1cm}. 
\end{equation}
The metric is then 
\begin{equation}
ds^2\,=\,{\cos}^2(au-bv)\,dx^2+{\cos}^2(au+bv)\,dy^2-2\,
\cos(au-bv)\cos(au+bv)\,dudv\hspace{1cm}
\end{equation}
and  it  can   be  checked  that  the electromagnetic   potential (52)
automatically satisfies the Maxwell's equations (51).  The solution is
well-defined for $-{\pi  / 2a}<u<{\pi /2a}$  and $-{\pi/  2b}<v<{\pi /
2b}$ and there is a curvature singulatity on the hypersurfaces $au \pm
bv={\pi /   2}$  as  it can  be  seen   by  calculating the  curvature
invariants            $R$,           $R_{\mu\nu}R^{\mu\nu}$        and
$R_{\mu\nu\lambda\kappa}R^{\mu\nu\lambda\kappa}$.  For example, $R$ is
given by 
\begin{equation}
	R\,=\,{2a     b  \sin (2au) \sin(2bv)\over   \cos^{3/2}(au-bv)
	\cos^{3/2}(au+bv)} 
\end{equation}
and diverges at   $au \pm bv={\pi /  2}$   including the 2-dimensional
boundaries $\{u={\pi/ 2a},v=0\}$ and $\{u=0,v={\pi/ 2b}\}$. 

	Introducing  the    following  Newmann-Penrose   null   tetrad
$(l,n,m,\bar{m})$ relatively to the space-time metric (56) 
\begin{eqnarray}
	l_{\mu} & =  &     \left(0,\cos(au-bv)\cos(au+bv),0,0\right)\\
	n_{\mu}  &    =  & \left(1,0,0,0  \right)\\   m_{\mu}    & = &
	\left(0,0,{1\over                 \sqrt{2}}\cos(au-bv),{i\over
	\sqrt{2}}\cos(au+bv)\right) 
\end{eqnarray} 
one finds the   following components   of the electromagnetic    field
associated with the potential (52) 
\begin{eqnarray}
	F_{\mu\nu} & = & {1\over \sqrt{2}}\left( \begin{array}{cccc} 0
	& 0 &  a e^{i \theta}  & a e^{-i  \theta} \\ 0  & 0  & -b e^{i
	\theta}f^{-1}&  -b e^{-i\theta}f^{-1}\\ -a  e^{i  \theta}  & b
	e^{i  \theta}f^{-1} &   0   &  0\\    -a e^{-i \theta}    &  b
	e^{-i\theta}f^{-1} & 0 & 0 \end{array} \right) 
\end{eqnarray}
where $f\equiv\cos(au-bv)\cos(au+bv)$.  The electromagnetic  invariant
and pseudo invariant are respectively given by 
\begin{eqnarray} 
F_{\mu\nu}F^{\mu\nu}     &        =   &    {4ab\sin(2\theta)     \over
\cos(au-bv)\cos(au+bv)}\nonumber\\     F_{\mu\nu} {^*F}^{\mu\nu} & = &
{4ab\cos(2\theta) \over \cos(au-bv)\cos(au+bv) } 
\end{eqnarray}
showing that the field is  null when $a$ or $b$  are zero and that the
electric and magnetic parts are parallel when $a=\pm b$.  Moreover one
sees that the electromagnetic field is like  the curvature singular ot
$au \pm bv={\pi/2}$. 

	The   null  hypersurfaces $u=const.$   are  generated by  null
geodesics with tangent  $\partial /\partial v$.  Their null generators
have expansion $\rho$ and shear $\sigma$ given by 
\begin{equation}
	\rho \,=\,{b\,\sin(2bv)\over 2\cos (au-bv)\cos (au+bv)}\,\,  ,
	\,\, \sigma\,=\,{b\,\sin(2au)\over 2\cos (au-bv)\cos (au+bv)} 
\end{equation}
One finds that  there is an  infinite focussing of the null congruence
at the  hypersurfaces $au \pm bv={\pi /2}$.   These null geodesics are
shear-free and  expansion-free when $b=0$ and  in this case correspond
to  plane   waves. The existence of these plane waves is a similar result 
as first obtained in general relativity in ref. \cite{Hogan} using 
Bertotti-Robinson space-times where the same formulas (63) hold. 
Furthermore, when    $b=0$,   the metric  (56)  is
conformally  flat. Similar properties  hold for the null hypersurfaces
$v=constant$ generated by the vector field $\partial /\partial u$ . 

	The properties of the space-time described  by the metric (56)
 are thus much different from   the Bertotti-Robinson solution for   the
 following reasons.     The Bertotti-  Robinson  solution   of general
 relativity is the only  solution of the Einstein-Maxwell theory which
 is   homogenous and has  a homogenous  non-null electromagnetic field
 \cite{EX} and moreover it is  conformally flat and without  curvature
 singularity.  Here, the electromagnetic field is also non-null but
 non homogenous.   Moreover, there is a  curvature singularity and the
 space-time  is  not  conformally  flat  when  both  $a$ and  $b$  are
 non-zero.


	Let us consider as in section  2, a null hypersurface $\Sigma$
along which are glued  two spacetimes $M^+$ and  $M^-$, $M^-$ being to
the  past of  $\Sigma$.  We  take  for $M^-$ the  vacuum solution of section 2
and  for $M^+$ the
above  solution.  The coordinates are  $(u,v_+,x_+,y_+)$ in $M^+$ with
$u>0$ and  $(u,v,x,y)$ in $M^-$ with $u<0$   -here again, we  drop the
minus   indices  for   any  quantity  refering  to    $M^-$.  The null
hypersurface  $\Sigma$ is located  at $u=0$ represents the propagation
of   a null  shell  and/or an  impulsive  gravitational   wave with an
electromagnetic shock wave.  The space-time metrics, the scalar fields
and the electromagnetic potentials only depend on the u-coordinate and
are  given  respectively  by  (31-32-34-39) with   (37)   in $M^-$ and
(52)-(56) with $b=0$ in $M^+$ {\it i.e.} 
\begin{equation}
ds_+^2\,=\,{\cos}^2 (au)\,(\,-2\,du dv^++dx_+^2+\,dy_+^2\,) 
\end{equation}
\begin{equation}
\varphi   _+\,   =  \,\sqrt   2   \,\log   \left\lbrack  {\cos   (au)}
\right\rbrack\, +\, {\varphi}_0^+ 
\end{equation}
\begin{equation}
	A_+\,=\,\sin          (au)(\cos\theta          dx_++\sin\theta
	dy_+)\hspace{1cm}. 
\end{equation}
As in section 2, the soldering of the two spacetimes along $\Sigma$ is
realized by making the identification 
\begin{equation}
(0,v_+,x_+,y_+) = (0,v - F(x,y),x,y)\hspace{1cm}, 
\end{equation} 
and   we take  $\xi^a   = (v,x,y)$  with  $a  =   1,2,3$, as intrinsic
parameters   on $\Sigma$,  $e_{(a)}=\partial/\partial\xi^a$ being  the
corresponding tangent basis  vectors and $n =  e_{(1)}$ the  normal to
$\Sigma$.  The  induced metrics on  $\Sigma$ from  $M^+$ and $M^-$ are
identical and equal to $\,g_{ab} = diag(0,1,1)$ and  one takes for the
``inverse''   (12)  $\,g^{ab}_* =  diag(0,1,1)$.  The  electromagnetic
potential vanishes on $\Sigma$.  The continuity of the scalar field on
$\Sigma$  implies $\varphi _0^+=\varphi_0$  and the  continuity of the
metric  requires that $U^{+}(0) = V^{+}(0)  = 0$  conditions which are
trivially   satisfied with our  choice.  A  convenient  choice for the
transversal $N$ corresponds to $N.n = -1$ and $N.e_{(2)} = N.e_{(3)} =
N.N = 0$. Although the  space-times $M^{\pm}$ are different from those
of section 2, one gets the same  components (40) and  the same is true
for the extrinsic curvature ${\cal K}_{ab}$: it  vanishes in $M^-$ and
its jumps $\gamma_{ab}$  is still given  by (41).  It follows that the
properties of the null shell  and the gravitational wave are unchanged
and characterized by the shift function $F(x,y)$.  For the jump in the
electromagnetic field across $\Sigma$, we get from (15) 
\begin{equation}
	\lambda   _\nu  \,=\,(0,0,a\cos\theta    , a\sin\theta       )
				\hspace{1cm}. 
\end{equation}
As $\lambda . n=0$, the null hypersurface  $\Sigma$ does not carry any
surface current.  Therefore,     it  is simply  the   history   of the
wave-fronts of a plane  electromagnetic  shock wave which  is entirely
characterized  by the  vector  ${\lambda}_{\nu}   $  and of  an  plane
impulsive gravitational  wave and at  the  same time  the history of a
null shell of matter. In the context of general relativity, it was first 
pointed out in \cite{Hogan} (using a system of coordinates continuous across 
the shell) that such a coexistence was possible in Bertotti-Robinson space-times. 

\begin{figure}
\epsffile{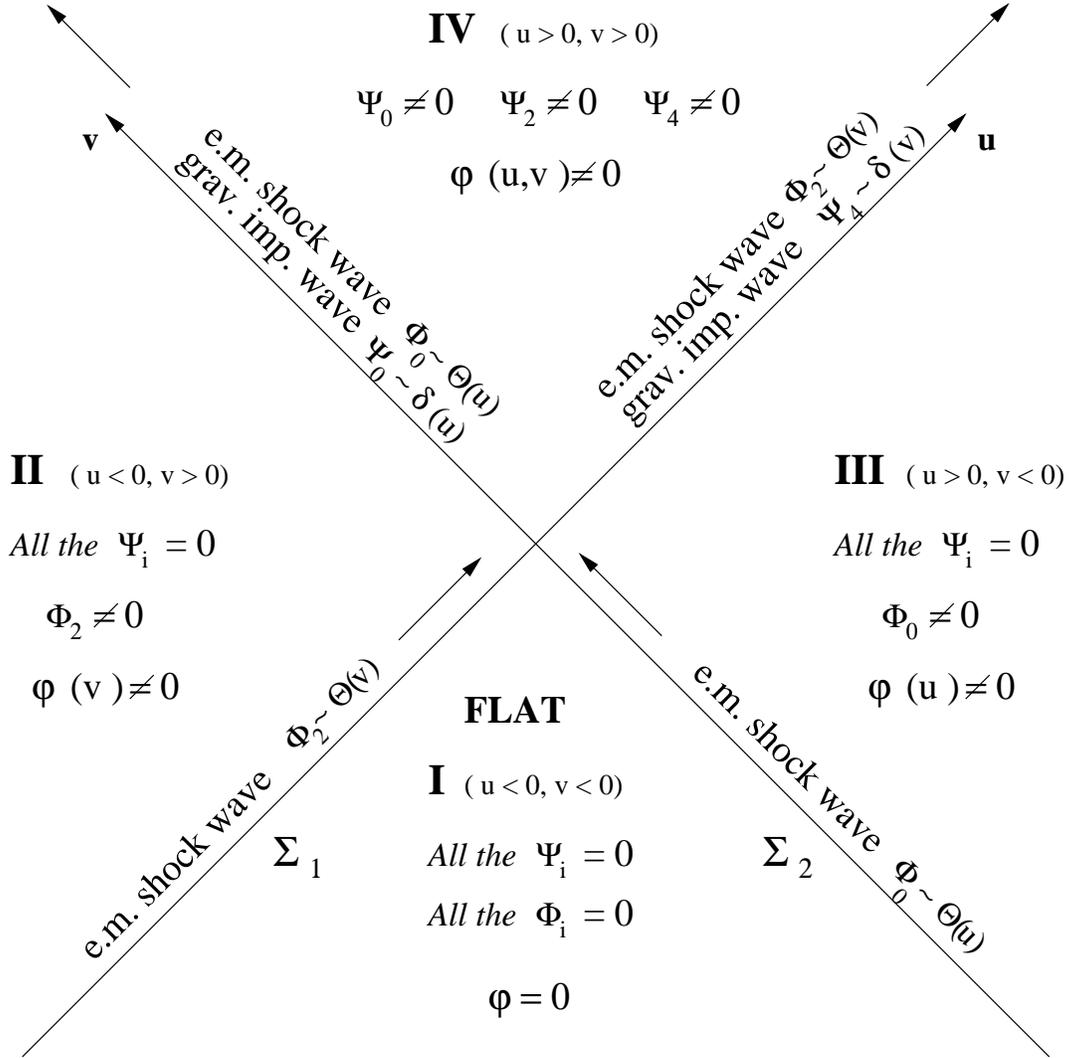} 
\caption{collision of two  plane  electromagnetic (e.m.)  shock  waves
giving two gravitational   impulsive (grav.imp.)  waves and  two plane
electromagnetic shock waves.} 
\label{collision}
\end{figure}

	Finally, the exact  solution (52-56) with  $a>0$ and $b>0$ can
be viewed as resulting  from  collision of two electromagnetic   shock
waves  approching  from     opposite  directions    and  with  aligned
polarization.  Let us divide the  space-time into four regions according
to fig. \ref{collision}. 
 
	In region I, the space-time is flat i.e.  the metric is of the
form (19) with $U=V=M=0$ and the scalar  field and the electromagnetic
potential are  zero.  In region II, the  metric, the  scalar field and
the electromagnetic field are given by 
\begin{equation}
	ds^2\,=\,{\cos}^2 (au)\,(\,-2\,du dv+dx^2+\,dy^2\,) 
\end{equation}
\begin{equation}
\varphi\, = \,\sqrt 2 \,\log \left\lbrack {\cos (au)} \right\rbrack 
\end{equation}
\begin{equation}
	A\,=\,\sin (au)(\cos\theta_1 dx+\sin\theta _1 dy)\hspace{1cm}. 
\end{equation}
and in region III 
\begin{equation}
	ds^2\,=\,{\cos}^2 (bv)\,(\,-2\,du dv+dx^2+\,dy^2\,) 
\end{equation}
\begin{equation}
\varphi\, = \,\sqrt 2 \,\log \left\lbrack {\cos (bv)} \right\rbrack 
\end{equation}
\begin{equation}
	A\,=\,\sin     (bv)(\cos\theta     _2      dx+\sin\theta    _2
	dy)\hspace{1cm}. 
\end{equation}
where  $\theta_1$ and     $\theta_2$ are   arbitrary constants.    The
space-time  regions  II and III   exhibit topological singularities at
$v={\pi/ 2b}$ and $u={\pi/ 2a}$ respectively.  These singularities are
not curvature singularities since  the curvature tensor remains finite
on  these hypersurfaces\footnotemark[1].  Moreover,  all the curvature
invariants       vanish          on    these        null   boundaries.
\footnotetext[1]{Actually,  all the  Riemann components  vanish except
$R_{uxux}=R_{uyuy}$ which have constant values $2 a^2$  and $2 b^2$ on
each boundary respectively.} 

Between the two regions I and II, the metric and  the scalar field are
$C^1$   and     the     electromagnetic   potential    $C^0$    across
$\Sigma_1\,=\,\{u=0\}$. Therefore, there is no impulsive gravitational
wave and no shell but an electromagnetic  shock wave propagating along
$\Sigma _1$.  In a similar way, between regions  I and III, the metric
and   the    scalar  field    are    $C^1$   across    the    boundary
$\Sigma_2\,=\,\{v=0\}$ and the electromagnetic potential is only $C^0$
and there is  only an  electromagnetic  shock wave  propagating  along
$\Sigma_2$. 

	The  space-time region  IV-see fig.\ref{collision}-  has the
metric (56), the electromagnetic  potential (52) and the  scalar field
(55).   One now  shows that this   choice in  region IV satisfies  the
boundary conditions for the re-attachement of the domains IV to II and
IV to  III   along the null   boundaries  $\Sigma _1$  and  $\Sigma_2$
respectively.  The metric,  the scalar  field and the  electromagnetic
potential are continuous across $\Sigma  _1$ between regions II and IV
and across $\Sigma _2$ between regions III and IV. But there is a jump
in the transverse derivatives   of the metric across $\Sigma_1$  which
proves  the existence of  an  impulsive  gravitational wave  which has
$\Sigma_1$ as history of its wave-fronts and propagating with an plane
electromagnetic shock  wave.  The   same  arguments applied    between
regions   III and  IV  lead   to the    conclusion that   an impulsive
gravitational wave and an electromagnetic  shock wave propagate  along
$\Sigma _2$ with this null hypersurface as their history. 

	The  interaction   of two electromagnetic    shock waves  then
produces   two gravitational    impulsive    waves propagating    with
electromagnatic  shock waves.  This is the  analog of te Bell-Szekeres
solution in general relativity but the interaction region now exhibits
a curvature singularity on the hypersurface  $au\pm bv={\pi/ 2}$ which
extends  back into the past  to the topological singularities $v={\pi/
2b}$ of the region II and $u={\pi/ 2a}$ in region  III and thus act as
``fold singularities'' \cite{JBG}.   In region IV, the non-vanishing 
Newmann-Penrose components of the electromagnetic field on the null
tetrad (58-60) are
	\begin{equation}
		\phi_0\,=\,-{a\over \sqrt 2}\,g \,\,\,,\,\,\, 
		\phi_2\,=\,-{b\over \sqrt 2\,f^2}\,g 
	\end{equation}
where $g\equiv {e^{i\theta}\over cos(au-bv)}-i\,{e^{i\theta}\over cos(au+bv)}$. 
A computation of the Weyl scalars
of the region IV reveals that this  region is not conformally flat. On
the   null  tetrad  (58-60),  the  non-vanishing  Newmann-Penrose Weyl
scalars are 
\begin{eqnarray}
	\Psi _0   &   =  &   b   \tan(au) \Theta(u)\delta(v)+{b^2\over
	2}(\tan^2 (au+bv)-\tan^2  (au-bv))\Theta(v) \\  \Psi _2 &  = &
	-{ab\over   6}{(\tan^2 (au+bv)-\tan^2 (au-bv))\over  \cos(au-bv)
	\cos(au+bv)}\,\Theta(u)\Theta(v)\\ \Psi _4  & =  & a  \tan(bv)
	\Theta(v)\delta(u)+{a^2\over      2}(\tan^2     (au+bv)-\tan^2
	(au-bv))\Theta(u) 
\end{eqnarray}
where $\Theta$ stands for the Heaviside  step function.  One concludes
that the region  IV   exhibits gravitational shock waves    which are
absent  in the Bell-Szekeres solution  where the four  regions are all
conformally  flat.   Moreover,  there  exists a ``Coulomb''  component
$\Psi_2$  in the  interaction region  which  can  be interpreted as  a
scattering effect of the two gravitational shock waves. 

\section*{Acknowledgments}
I am  grateful to  Claude Barrab\`es  for  his generous assistance and
endless encouragement and I especially thank P.A. Hogan for communicating 
prior to publication his results \cite{Hogan} on plane waves in Bertotti-Robinson 
space-times and for many helpful discussions.

%
%

%
%

\setcounter{equation}{0}
\renewcommand{\theequation}{A.\arabic{equation}} 
\section*{\centerline{Appendix A}}
\appendix 

When the metric $g_{\mu\nu}$ has the Szekeres form (19) and the scalar
field $\varphi$ only depends on the null  coordinates $u,v$, the field
equations (1)-(2) reduce to 
\begin{eqnarray}
R_{uu}  &    =     &   U_{uu}+U_u   M_u-{1\over     2}(U_u^2+V_u^2)\,=
		\,8\pi\,T_{00}+2\,({\partial}_{u}\varphi)^2\\ R_{vv} &
		=   &U_{vv}+U_v     M_v-{1\over     2}(U_v^2+V_v^2)\,=
		\,8\pi\,T_{11}+2\,({\partial}_{v}\varphi)^2\\ R_{uv} &
		=  &  M_{uv}+U_{uv}-{1\over   2}(U_v  U_u+V_u   V_v)\,
		=\,2\,{\partial}_{u}\varphi  \,{\partial}_{v}\varphi\\
		R_{xx}  &   =  &   e^{M-U+V}\,\left\lbrace  V_{uv}+U_u
		U_v-U_{uv}    -{1\over             2}(U_u      V_v+U_v
		V_u)\right\rbrace\,=\,8\pi\,T_{22}\\  R_{yy} &   =   &
		e^{M-U-V}\,\left\lbrace     -V_{uv}+U_u     U_v-U_{uv}
		+{1\over             2}(U_u                    V_v+U_v
		V_u)\right\rbrace\,=\,8\pi\,T_{33} 
\end{eqnarray}
and 
\begin{equation}
\Box         \varphi          \,                =                   \,
			e^M\,(-2\,{\partial}_{uv}\varphi+U_v\,{\partial}_u\varphi
			+U_v\,{\partial}_u\varphi)\,=\,-4\pi \alpha T 
\end{equation}
where $T=g^{\mu\nu}T_{\mu\nu}$.  In this space-time, the stress-energy
tensor $T_{\mu\nu}$ of the  electromagnetic field corresponding to the
potential (52) of the form $A=\lambda dx+\mu dy$,  is diagonal and its
components are given by 
\begin{eqnarray}
4\pi\,T_{00} & = & e^{U-V}({{\lambda}_u}^2+e^{2V}{{\mu}_u}^2)\\
4\pi\,T_{11} & = & e^{U+V}({{\lambda}_v}^2+e^{2V}{{\mu}_v}^2)\\
4\pi\,T_{22} & = &-e^M \, ({\lambda}_u{\lambda}_v-e^{2V}{\mu}_u{\mu}_v)\\
4\pi\,T_{33} & = &-e^M \, ({\mu}_u{\mu}_v-e^{-2V}{\lambda}_u{\lambda}_v)
\end{eqnarray}

\end{document}